\documentclass[aps,prb,groupedaddress,preprintnumbers,amsmath,amssymb,twocolumn,notitlepage,10pt]{revtex4-1}
\usepackage{graphicx}
\usepackage{dcolumn}
\usepackage{bm}
\usepackage[T1]{fontenc}
\usepackage[latin1]{inputenc}
\usepackage[english]{babel}
\usepackage{amsmath}
\usepackage{amssymb}
\usepackage{amstext}
\usepackage{braket}
\usepackage{subfigure}
\usepackage[rightcaption]{sidecap}

\newcommand{\bx}{\mathbf{x}}
\newcommand{\bb}{\mathbf{b}}
\newcommand{\be}{\mathbf{e}}

\newcommand{\mean}[1]{\left < #1 \right >}
\begin{document}
\title{Multi-Component Quantum Gases in Spin-Dependent Hexagonal Lattices}
\author{P. Soltan-Panahi}
\affiliation{Institut f\"ur Laser-Physik, Universit\"at Hamburg, Luruper Chaussee 149, 22761 Hamburg, Germany}
\author{J. Struck}
\affiliation{Institut f\"ur Laser-Physik, Universit\"at Hamburg, Luruper Chaussee 149, 22761 Hamburg, Germany}
\author{P. Hauke}
\affiliation{
    ICFO -- Institut de Ci\`encies Fot\`oniques, Mediterranean Technology Park, E-08860 Castelldefels (Barcelona), Spain
    }
\author{A. Bick}
\affiliation{Institut f\"ur Laser-Physik, Universit\"at Hamburg, Luruper Chaussee 149, 22761 Hamburg, Germany}
\author{W. Plenkers}
\affiliation{Institut f\"ur Laser-Physik, Universit\"at Hamburg, Luruper Chaussee 149, 22761 Hamburg, Germany}
\author{G. Meineke}
\affiliation{Institut f\"ur Laser-Physik, Universit\"at Hamburg, Luruper Chaussee 149, 22761 Hamburg, Germany}
\author{C. Becker}
\affiliation{Institut f\"ur Laser-Physik, Universit\"at Hamburg, Luruper Chaussee 149, 22761 Hamburg, Germany}
\author{P. Windpassinger}
\affiliation{Institut f\"ur Laser-Physik, Universit\"at Hamburg, Luruper Chaussee 149, 22761 Hamburg, Germany}
\author{M. Lewenstein}
\affiliation{
    ICFO -- Institut de Ci\`encies Fot\`oniques, Mediterranean Technology Park, E-08860 Castelldefels (Barcelona), Spain
    }
\author{K. Sengstock}
\email{sengstock@physik.uni-hamburg.de}
\affiliation{Institut f\"ur Laser-Physik, Universit\"at Hamburg, Luruper Chaussee 149, 22761 Hamburg, Germany}

\begin{abstract}
Periodicity is one of the most fundamental structural characteristics of systems occurring in nature. The properties of these systems depend strongly on the symmetry of the underlying periodic structure. In solid state materials -- for example -- the static and transport properties as well as the magnetic and electronic characteristics are  crucially influenced by the crystal symmetry. In this context, hexagonal structures play an extremely important role and lead to novel physics like that of carbon nanotubes or graphene. Here we report on the first realization of ultracold atoms in a spin-dependent optical lattice with hexagonal symmetry. We show how combined effects of the lattice and interactions between atoms lead to a forced antiferromagnetic N\'eel order when two spin-components localize at different lattice sites. We also demonstrate that the coexistence of two components -- one Mott-insulating and the other one superfluid -- leads to the formation of a forced supersolid. Our observations are consistent with theoretical predictions using Gutzwiller mean-field theory.
\end{abstract}
\maketitle
In recent years, ultracold atoms in  periodic potentials\cite{Jaksch1998,Greiner2002} have been widely recognized as an important tool to simulate solid state systems and study their transport\cite{Trotzky2008a} and magnetic properties\cite{Lewenstein2007, Bloch2008}. Different types of spin- and/or state-dependent lattices have been implemented and studied\cite{Jaksch1999,Mandel2003,Lee2007,Lin2009,Lin2009b,McKay2009}. So far, most experiments with ultracold quantum gases have been carried out in lattices of cubic symmetry. However, recent theoretical developments\cite{Eckardt2010,Hofstetter2007,Lee2009} are aiming at systems with a hexagonal geometry. In particular, carbon nanotubes\cite{Geim2007}, graphene\cite{Iijima1991} and a large number of other carbon-based compounds show fascinating effects and exhibit particularly rich quantum phases\cite{Du2009}. In this paper we discuss the first realization of ultracold quantum gases in a hexagonal, spin-dependent optical potential and demonstrate how it can be used to tailor quantum phases of spin-mixtures and their dynamics. We show that the combination of interactions between different spin-states and the spin-dependent lattice potential leads to novel quantum phases: a forced antiferromagnetic N\'eel order when two spin-components localize in different sublattices, and a \textit{forced} supersolid phase, with one spin-component being localized in one sublattice and inducing a density modulation on the other superfluid spin-component. These phenomena are studied by exploiting a novel technique of state- and site-selective microwave-spectroscopy. Examining the impact on the onset of the superfluid to Mott-insulator transition of the respective spin-state, we demonstrate furthermore that the mobility of particles in the lattice can be adjusted by immersing a well localized spin-crystal into a superfluid bath. We calculate the corresponding phase diagrams within Gutzwiller mean-field theory and find good agreement with the experimental results.
\section{Generation of a hexagonal, spin-dependent optical lattice}
\begin{figure*}
\includegraphics[width=0.99\textwidth]{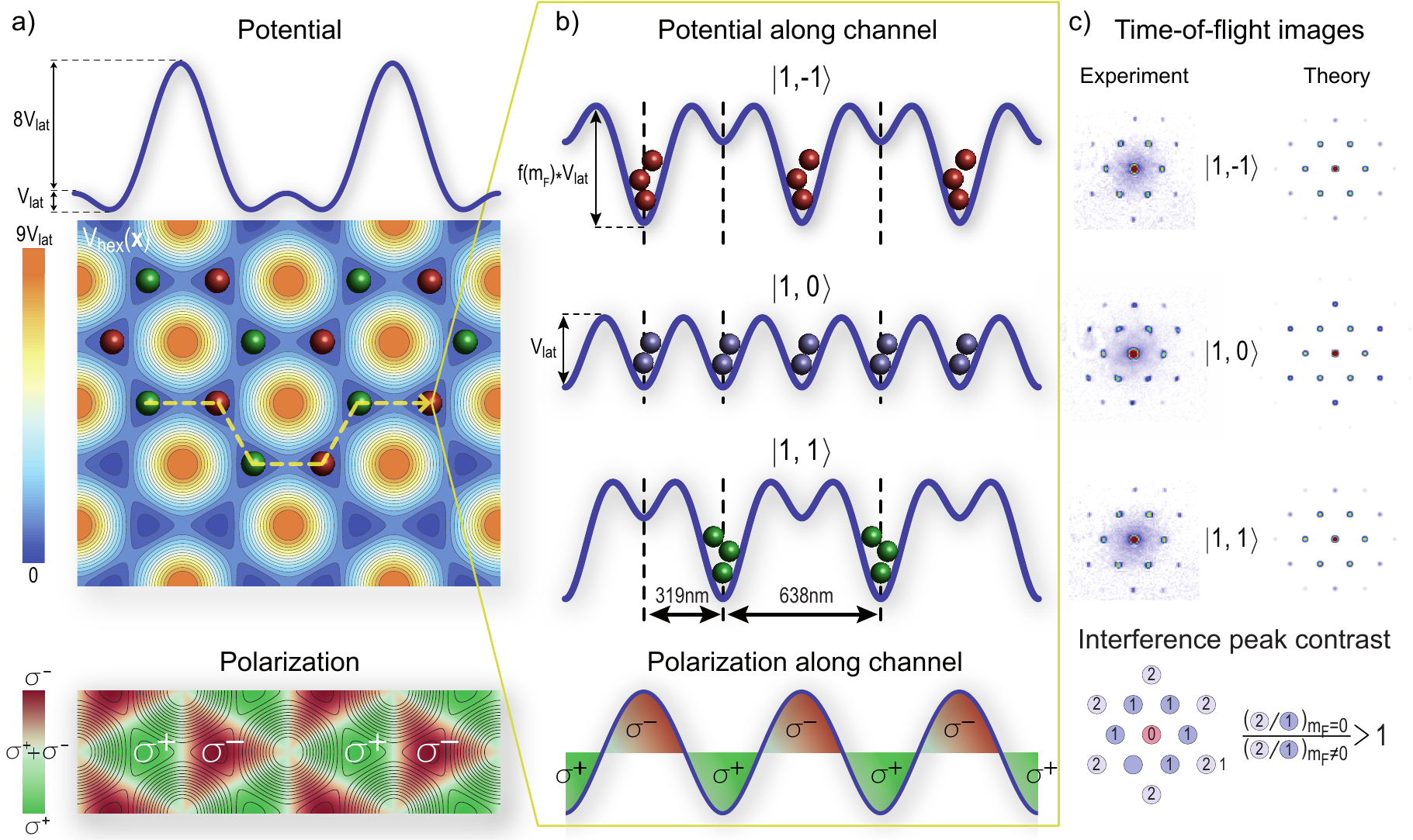}
\caption{
        \textbf{a} Hexagonal optical lattice with alternating $\sigma^+$ (green bullets) and $\sigma^-$ (red bullets) polarization structure. The upper 1D plot shows a cut through the 2D potential. The tunneling of atoms loaded into the potential is mainly restricted to channels like the one indicated by the orange dashed line. The lowest graph shows the 2D polarization distribution in the lattice ranging from fully $\sigma^+$(green) to $\sigma^-$(red) polarized. The hexagonal structure is based on two triangular sublattices which are translated with respect to each other.
        \textbf{b} One-dimensional potential structure along the channel shown in subfigure \textbf{a} for particles in different Zeeman states $\ket{F,m_F}$. The modulation depth $f(m_F)V_{\mathrm{lat}}$ of the channel depends on the $m_F$-state and is given by \{1,1.7,2.5\}$V_{\mathrm{lat}}$ for $m_F=\{0,\pm 1,\pm 2\}$ and $\lambda=830\mathrm{nm}$. The degeneracy of neighboring lattice sites for atomic substates with $m_F\neq 0$ is lifted by the polarization- and atomic state-dependence of the ac Stark shift. The lower part shows the corresponding light polarization. When the potential is deep enough, atoms in the different substates will localize as indicated.
        \textbf{c} Time-of-flight images of superfluid samples prepared in different Zeeman substates released from the lattice (left) and corresponding calculated quasi-momentum distributions (right). The second- to first- order interference peak ratio changes from the \textit{triangular} to the \textit{hexagonal} lattice occupation which is in direct agreement with the different structure factors of the two lattice geometries.}
\label{fig:1}%
\end{figure*}

The basic structure of the spin-dependent hexagonal optical lattice discussed in this paper is illustrated in Fig.~\ref{fig:1}. Three laser-beams intersecting under an angle of $120^{\circ}$ and with each beam linearly polarized in the plane of intersection lead to the formation of local potential minima in a hexagonal structure. For neighboring sites along the vertices of the hexagonal lattice as for example indicated in Fig.~\ref{fig:1}a by a dashed line, the local polarization alternates between $\sigma^+$ and $\sigma^-$. As atoms in a light field experience a polarization-dependent ac Stark shift, the potential at these $\sigma^+$ and $\sigma^-$ sites is different for different atomic Zeeman substates labeled by $m_F$. The potential can be written as:
\begin{equation}
  V(\bx) = V_{\mathrm{hex}}\left(\bx\right) + m_F g_F \mu_{\mathrm{B}} B_{\mathrm{eff}}\left(\bx\right), \label{potential}
\end{equation}
where the polarization of the light field $\mathcal{P}\left(\bx\right)$ ($+1$ for pure $\sigma^+$ and $-1$ for pure $\sigma^-$ polarizations) is mapped onto a pseudo-magnetic field $B_{\mathrm{eff}}\left(\bx\right)\propto -V_{\mathrm{hex}}\left(\bx\right) \mathcal{P}\left(\bx\right)/\mu_{\mathrm{B}}$, $g_F$ the Land\'e $g$-factor and $\mu_{\mathrm{B}}$ the Bohr magneton (for details see appendix). The potential consists of a spin-independent part $V_{\mathrm{hex}}\left(\bx\right)$ of hexagonal symmetry and a state-dependent super-lattice emerging from the local pseudo-magnetic field $B_{\mathrm{eff}}\left(\bx\right)$. According to the local polarization, we denote the emerging triangular substructures as $\sigma^+$ and $\sigma^-$ lattice (see Fig.~\ref{fig:1}a). The hexagonal lattice can therefore also be regarded as a triangular lattice with a bi-atomic basis where the atoms occupy $\sigma^+$ and $\sigma^-$ sites as indicated in Fig.~\ref{fig:1}a by green and red bullets.
For typical experimental parameters, the mobility of the atoms is predominantly restricted to the hexagonal channel structure connecting the local minima of $V_{\mathrm{hex}}\left(\bx\right)$. One possible tunneling channel is indicated in Fig.~\ref{fig:1}a by a dashed line. Note that the large central peak maxima ($9 V_{\mathrm{lat}}$) prevent a direct diagonal tunneling through these maxima.
Figure~\ref{fig:1}b illustrates the different lattice potentials along a tunneling channel for the $F=1$ hyperfine ground-state manifold of $^{87}$Rb. The $F=2$ hyperfine states exhibit a similar behavior with twice the modulation amplitude for $\left|m_F\right|=2$ (compare Eq.~(\ref{potential})).
\\
To study atomic quantum phases in the lattice, we load an ensemble of ultracold $^{87}$Rb ground-state atoms into the lattice. This consists of either pure or specific compositions of different hyperfine- and magnetic Zeeman-states. Due to the additional spin-dependent spatial variation of the potential, atoms with $m_F\neq0$ are preferably confined to either the $\sigma^+$ or the $\sigma^-$ triangular lattice structure. In contrast, atoms in the $m_F=0$ state will distribute homogeneously over both the $\sigma^+$ and $\sigma^-$ lattices (Fig.~\ref{fig:1}b).  For convenience, we refer to these different configurations as {\it triangular} ($m_F\neq0$) and {\it hexagonal} states ($m_F=0$). This spin-dependent spatial distribution determines the structure factor of the lattice, which in turn determines the quasi-momentum distribution of the atoms. We experimentally observe this difference in the structure factors by mapping the quasi-momentum spectrum to the spatial density distribution of the ensemble via time-of-flight images. Typical results for \textit{triangular} and \textit{hexagonal} initial states are shown in Fig.~\ref{fig:1}c together with the corresponding theoretical prediction from a one-particle band structure calculation. The observed second- to first-order interference peak ratio (see Fig.~\ref{fig:1}c bottom) is significantly larger for the {\it hexagonal} states as compared to {\it triangular} configurations. It typically ranges between two and four for our experimental parameters, which is well reproduced by calculations of the Bloch functions and their quasi-momentum distribution for the corresponding spin-dependent lattice geometry.
\begin{figure*}
\includegraphics[width=0.99\textwidth]{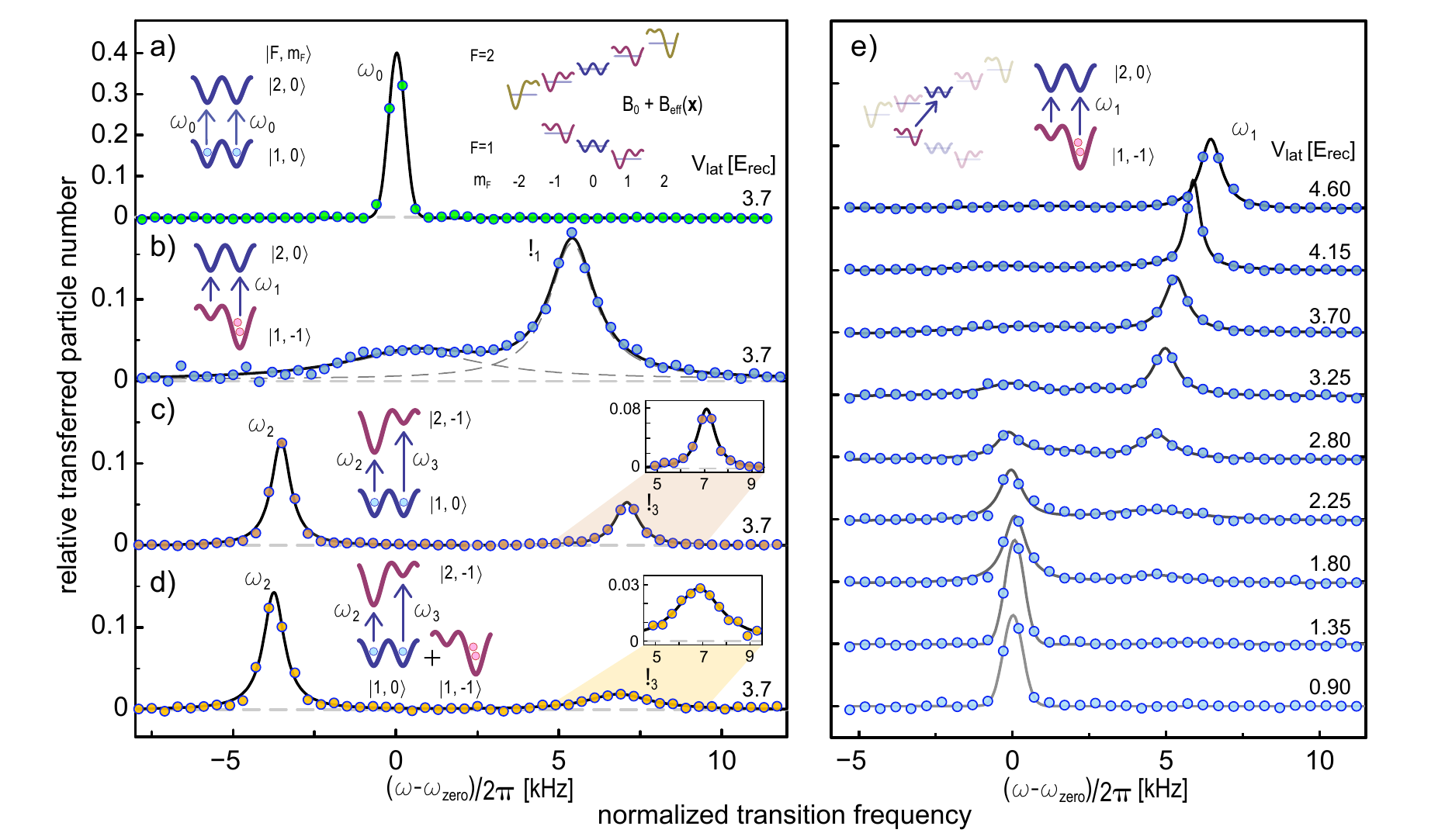}
        \caption{Results of microwave-spectroscopy measurements. The addressed transitions are indicated in the insets. For comparison, the lattice depth $V_{\mathrm{lat}}$ is given for each data set.
        Graphs \textbf{a}-\textbf{c},\textbf{e} show measurements for single-component samples and graph \textbf{d} for a spin-mixture loaded into the lattice. The frequency axis is shifted by the resonance frequency $\omega_{\mathrm{zero}}$ of the respective sample for vanishing 2D-lattice ($V_{\mathrm{lat}}\!=\!0$).
        \textbf{a} Transferring the atoms from a \textit{hexagonal} state $\left |1,0\right>$ to another \textit{hexagonal} state $\left |2,0\right>$ leads to a single Fourier limited transition line at $\omega_0$.
        \textbf{b} Spectroscopy of the {\it triangular} state $\ket{1,-1}$ to $\ket{2,0}$. For sufficiently deep lattice potentials, the atoms start to localize in the deeper wells. The transition spectrum therefore shows a dominant resonance at the corresponding frequency $\omega_1$.
        \textbf{c} Addressing the $\ket{1,0} \rightarrow \ket{2,-1}$ transition leads to two well separated resonance lines $\omega_2$ and $\omega_3$ corresponding to the energy splitting of the potential wells due to the pseudo-magnetic field $B_{\mathrm{eff}}$. The integrals of the individual peaks are proportional to the number of atoms in the respective wells. The different height of the peaks reflects the different Franck-Condon overlap between the initial and the different final states.
        \textbf{d} Interaction induced localization in spin-mixtures. An equal mixture of $\ket{1,0}$ and $\ket{1,-1}$ atoms is prepared and the occupation in the lattice wells is determined for the $\ket{1,0}$ atoms. The presence of $\ket{1,-1}$ atoms suppresses the population of the $\ket{1,0}$ atoms at $\sigma^-$ sites and only the $\sigma^+$ resonance remains.
        \textbf{e} Spectroscopy data for different lattice depths. The peak centered at 0 corresponds to the transition frequency in the system without lattice. For increasing lattice depth, the atoms redistribute and localize in the deeper wells.}
        \label{fig:microwavespectroscopy}%
\end{figure*}
\section{State- and site-selective microwave spectroscopy}
To obtain a detailed understanding of the rich physics arising from different lattice occupations we have developed a novel method for state- and site-selective microwave spectroscopy. It allows for {\it in-situ} investigations of the spatial spin-ordering in the hexagonal and triangular lattice structures. It is suitable for studying different regimes ranging from the superfluid to the Mott-insulating phase. Previously, a different kind of polarization-dependent radio-frequency spectroscopy method was used to study cubic lattices\cite{Lee2007}. Our method is based on the spatially varying transition frequencies between populated $m_F$-states to initially non-occupied $m_{F'}$-states.
The potential energy difference between the initial and the final state is given by $\Delta V(\bx) = \mu_{\mathrm{B}} \left(m_{F'}+m_F\right)\left(B_0+B_{\mathrm{eff}}\left(\bx\right)\right)/2+E_{\mathrm{hfs}} $,
where $B_0$ is an additionally applied homogeneous magnetic guiding field and $E_{\mathrm{hfs}}$ is the $^{87}$Rb-hyperfine splitting.
The spectroscopy signal is obtained by detecting and normalizing the number of atoms in the respective $m_F$- and $m_{F'}$-states after applying a microwave pulse, then releasing the sample from the lattice and performing Stern-Gerlach separation.
In the following we discuss central results of the microwave spectroscopy method, first for single component samples, then for spin-mixtures. Typical data for both scenarios is shown in Fig.~\ref{fig:microwavespectroscopy}.
\\[0.4cm] 
The spatial variation of the differential potential shift $\Delta V(\bx)$ vanishes for transitions with $m_{F'}+m_F=0$. Thus a single component sample initialized in the  $\ket{F,m_F}=\ket{1,0}$ state shows a single transition line when the  state $\ket{2,0}$ is addressed (Fig.~\ref{fig:microwavespectroscopy}a). When transferring atoms from the {\it hexagonal} state $\ket{1,0}$ to a {\it triangular} final state ($m_{F'}+m_F\neq0$; the data in Fig.~\ref{fig:microwavespectroscopy}c shows $\ket{1,0}$ to $\ket{2,-1}$ for $V_{\mathrm{lat}}=3.7 E_{\mathrm{rec}}$), the spectrum splits up into two lines according to the energy  difference between the neighboring lattice sites. Here, the left (right) transition peak refers to the transition from the first band of the {\it hexagonal} state $\ket{1,0}$ to the first (second) band of the {\it triangular} state $\ket{2,-1}$ where the atoms are mostly confined to the $\sigma^{-}(\sigma^{+})$ sites. We measure the two transition frequencies $\omega^{\mathrm{exp}}_2-\omega_{\mathrm{zero}}=2\pi\!\times\!(-3.6\pm 0.2)$kHz and $\omega^{\mathrm{exp}}_3-\omega_{\mathrm{zero}}=2\pi\!\times\!(7.1\pm 0.2)$kHz, where $\omega_{\mathrm{zero}}$ is the resonance frequency without 2D-lattice. We find excellent agreement with the theoretical expectations of $\omega^{\mathrm{theo}}_2-\omega_{\mathrm{zero}}=-2\pi\!\times \!3.8\,$kHz and $\omega^{\mathrm{theo}}_3-\omega_{\mathrm{zero}}=2\pi\!\times\!7.1 \,$kHz obtained using an {\it ab initio} single-particle 2D band-structure calculation. Note that for the hexagonal lattice the Bloch functions do not separate like it is the case for cubic lattice.
\\[0.4cm]
Moreover, we compare the particle transfer efficiencies of the two transitions with the calculated predictions. Experimentally, the transfer efficiencies are determined evaluating the integrals of the corresponding transition peaks. For the measurement shown in Fig.~\ref{fig:microwavespectroscopy}c we measure the transfer efficiency of the $\omega_2$-transition to be $(2.0 \pm 0.2)$ times larger than for the $\omega_3$-transition. This is in very good agreement with the calculated ratio of the Franck-Condon overlaps between the initial state and the two final states resulting in a $1.9$ times larger transfer efficiency for the $\omega_2$-transition. It also confirms the expected homogeneous density distribution of $m_F\!=\!0$ atoms at $\sigma^+$ and $\sigma^-$ sites. Therefore the spectroscopy allows to measure the relative occupations of the two sublattices $\sigma^+$ and $\sigma^-$. It can equally well be applied to study other initial spin-states e.g.~a state of $\ket{1,-1}$ atoms. Figures \ref{fig:microwavespectroscopy}b and e show the $\ket{1,-1}$ to $\ket{2,0}$ transition amplitudes for different lattice depths and clearly reveal the exclusive occupation of the $\sigma^-$ sites for sufficiently deep lattices. For shallow lattices, the microwave transition frequency coincides with the one for a vanishing lattice potential. A crossover can be observed at intermediate lattice depths. The observed shift of $5-7\mathrm{kHz}$ for deeper lattices is reproduced by calculations within a harmonic approximation, whereas the signals for shallow and intermediate lattices cannot easily be explained in the single-site approximation. When initializing the atoms in $\ket{1,+1}$, we observe the equivalent signals corresponding to the occupation of the $\sigma^+$ sites and thus demonstrate the ability to actively force the generation of a magnetically ordered Ne\'el state with predefined magnetization by using a mixture of different spin-states, e.g., $\ket{1,-1}$ with $\ket{1,+1}$.
\\[0.4cm]
The experimental setup allows for the preparation of arbitrary spin-mixtures in the spin-dependent optical lattice. The interplay of the interaction between different spin-states and the lattice potential has a crucial impact on the static and dynamic properties of the system. Particularly interesting is the case of a mixture of $\ket{1,0}$ and $\ket{1,-1}$ atoms. While the $\ket{1,-1}$ atoms tend to localize in the $\sigma^-$ sites, the $\ket{1,0}$ atoms fill the hexagonal lattice homogeneously as it was shown above for pure atomic samples (Figs.~\ref{fig:microwavespectroscopy}a-c, e). In case of a mixture of these spin-components, the additional repulsive {\it inter-}species interaction causes the $\ket{1,-1}$ atoms to imprint a periodic density modulation on the $\ket{1,0}$ atoms. Experimentally we observe an almost vanishing transition amplitude of the peak corresponding to the $\sigma^-$ sites as shown in Figure \ref{fig:microwavespectroscopy}d. Comparison of the integrated overall transition probability with the one obtained for pure systems indicate that ($30\pm 5$)\% of the $m_F=0$ atoms, initially sitting at $\sigma^-$ sites, have been transferred to $\sigma^+$ sites. This crystalline order imprinted onto the $\ket{1,0}$ atoms by the $\ket{1,-1}$ atoms can appear in two different scenarios: First, when the $\ket{1,0}$ atoms are localized on the $\sigma^+$ sites, the system forms an alternating Mott-insulator. In this case, the mixture resembles a localized alternating spin-ordering. Second, when the $\ket{1,0}$ atoms are still superfluid the system shows an off-diagonal long-range order. Together with the spatial modulation due to the interaction with the other spin-component, this indicates a supersolid like behavior\cite{Buechler2003,Titvinidze2008}. By varying the depth of the lattice both scenarios have been realized and indeed, we experimentally observe that the $\ket{1,0}$ atoms were driven across the superfluid-to-Mott-insulator transition referring to both cases.
\begin{figure}
\includegraphics[width=0.35\textwidth]{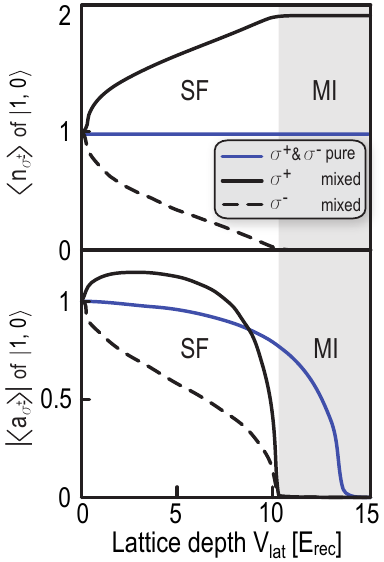}
    \caption{
    Results of Gutzwiller mean-field theory for \textbf{a}) occupation number $\mean{\hat n_{\sigma^{\pm}}}$ and \textbf{b}) absolute value of the superfluid order parameter $|\mean{\hat a_{\sigma^{\pm}}}|$ of $\ket{1,0}$ atoms. Two different ensembles are considered; a pure ensemble of $\ket{1,0}$ atoms (blue line) and a 50:50 mixture with additional $\ket{1,-1}$ atoms (black lines). The solid (dashed) black line is the value at $\sigma^+$ ($\sigma^-$) sites. In a pure system the values are identical at $\sigma^+$ and $\sigma^-$ sites (blue line). The particle number of each spin-state is fixed to two per unit cell.}
    \label{fig:SS}%
\end{figure}
\\[0.4cm]
\noindent
Our observations are strongly supported by a theoretical analysis using the mean-field Gutzwiller approximation\cite{Lewenstein2007} (see Appendix). In Fig.~\ref{fig:SS} we demonstrate that an equal mixture of $\ket{1,0}$ and $\ket{1,-1}$ atoms results in a \emph{forced supersolid}. This is shown with the aid of the absolute value of the superfluid order parameter $\mean{\hat a_{\sigma^{\pm}}}$ (top) and the occupation number $\mean{\hat n_{\sigma^{\pm}}}$ (bottom) of the $\ket{1,0}$ state at $\sigma^+$ and $\sigma^-$ sites assuming a filling of two particles per unit cell. For comparison, we also include the calculations for a pure $\ket{1,0}$ system. Here, $|\mean{\hat a_{\sigma^+}}|$ and $|\mean{\hat a_{\sigma^-}}|$ as well as $\mean{\hat n_{\sigma^+}}$ and $\mean{\hat n_{\sigma^-}}$ coincide for all $V_{\mathrm{lat}}$ revealing the expected homogeneous atom distribution within the lattice. This changes when $\ket{1,-1}$ atoms are admixed to the $\ket{1,0}$ atoms. At already moderate lattice depths the $\ket{1,-1}$ particles are localized at the $\sigma^-$ sites. As a consequence the $\ket{1,0}$ component is repelled from these sites and the corresponding occupation number and superfluid order parameter become spatially modulated, which is characteristic for a supersolid. Moreover, we observe that the effective blocking of $\sigma^-$ sites by $\ket{1,-1}$ particles forces the $\ket{1,0}$ atoms to undergo the transition to the Mott--insulator already at weaker lattice depths than in the pure case. Hence an admixture of a second spin-component allows us to tune the critical point of the SF--MI transition as well as the spatial density modulation and superfluid order parameter.
\begin{figure*}
\includegraphics[width=.99\textwidth]{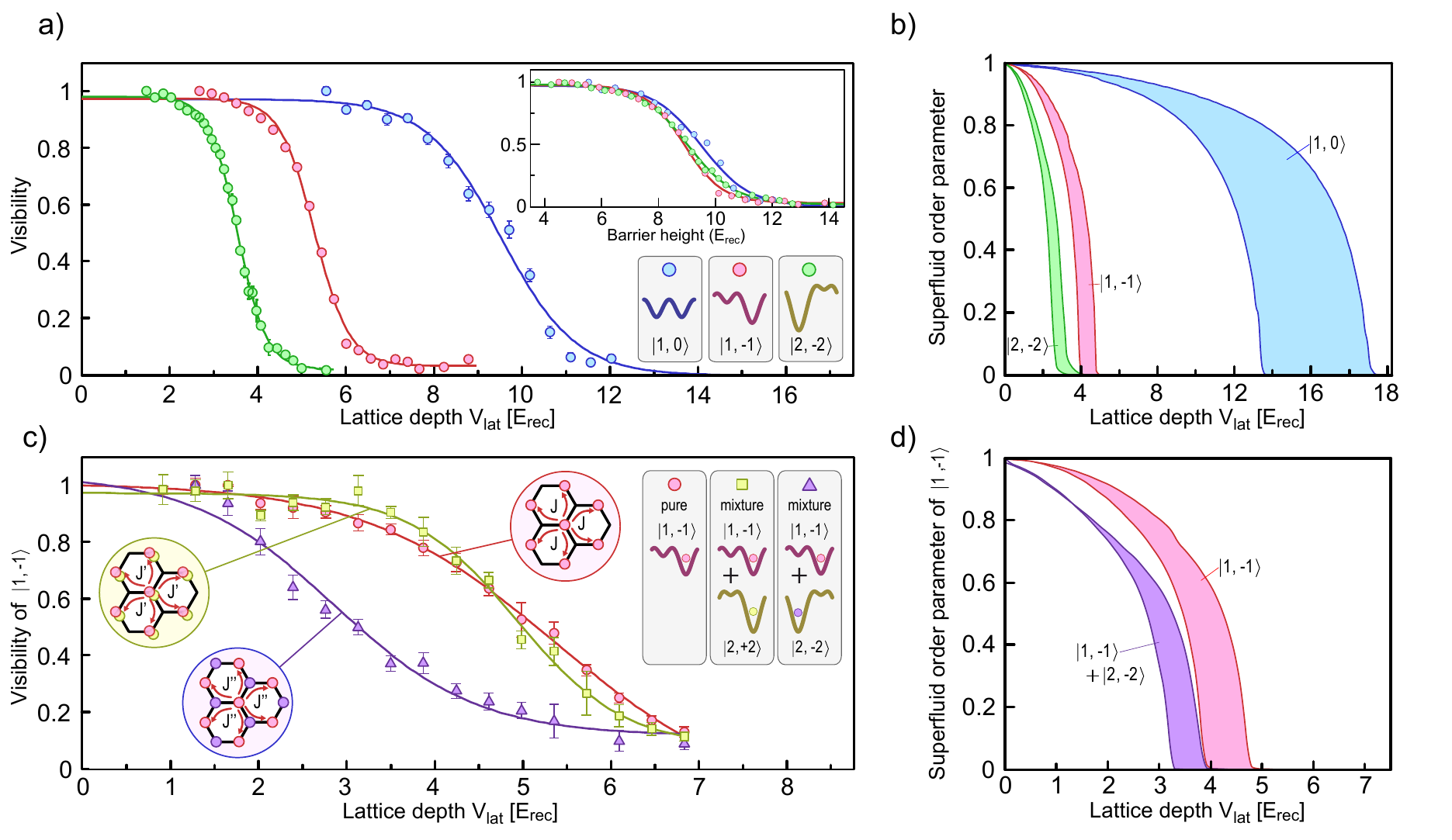}
        \caption{Experimental observation of the superfluid to Mott-insulator (SF--MI) transition for pure and mixed atomic spin-state system averaged over typically 7-10 experimental runs (\textbf{a} and \textbf{c}) and the corresponding theoretical superfluid order parameter normalized to the square root of the total occupation number for the same configurations, calculated within a mean-field Gutzwiller approach (\textbf{b} and \textbf{d}).
        \textbf{a} Visibility curves for different single spin-components of the optical lattice as function of the lattice depth $V_{\mathrm{lat}}$, which is proportional to the lattice laser beam powers. In the inset, the data is presented with the lattice depth renormalized to the height of the nearest neighbor tunneling barrier.
        \textbf{b} Theoretical calculation of the superfluid order parameter normalized to the square root of the total occupation number for the experimental settings presented in \textbf{a}. The approach of the superfluid order parameter to zero indicates the position of the SF--MI transition. The areas represent the results for lattice filling factors between two and four per unit cell as expected for our experimental parameters.
        \textbf{c} Comparison of visibility curves for pure $\ket{1,-1}$ and 50:50 mixtures of $\ket{1,-1}$ with $\ket{2,-2}$ ($\ket{2,+2}$) atoms. The insets indicate the preferred probability distributions of the atoms for the different mixtures and the modification of the tunneling matrix element $J$ of $\ket{1,-1}$.
        \textbf{d} Theoretical calculation of the superfluid order parameter of the  $\ket{1,-1}$ state for filling factors between two and four per unit cell.}
        \label{fig:3}%
\end{figure*}
\section{Superfluid to Mott-insulator transition in a spin-dependent optical lattice}
Transport properties play an important role in understanding, for example, the conductivity of solid-state systems. For the quantum optical counterparts of such systems, the optical lattices, these properties are essentially governed by the interplay of the on-site interaction $U$ with the the tunnel matrix element $J$ and are usually well described by the Bose--Hubbard model\cite{Jaksch1998} (see Appendix).
In optical lattice experiments, the ratio of $J/U$ is directly adjustable through the power of the lattice beams allowing to drive the ensemble from the superfluid to the Mott-insulating state. We use the interference pattern contrast (visibility) as an indicator for this transition\cite{Gerbier2005}. Results of such measurements are shown in Figure~\ref{fig:3}a for different single-component samples. We compare the SF--MI transition of a \textit{hexagonal} state $\ket{1,0}$ with that of the \textit{triangular} states $\ket{1,-1}$ and  $\ket{2,-2}$ and observe that the interference contrast curves are considerably shifted with respect to one another.\\
We start by analyzing the situation for the $\ket{1,-1}$ and the $\ket{2,-2}$ states. As these states experience an additional modulation of the hexagonal potential caused by $B_{\mathrm{eff}}(\bx)$, resonant tunneling is only possible to the six  {\it next-nearest neighbor sites} in the lattice. Due to this additional modulation, the tunneling barrier between the occupied lattice sites increases with $|m_F|$ (compare Eq.~(\ref{potential})). The SF--MI transition is thus expected to occur at lower overall beam powers for increasing values of $|m_F|$. However, the two transitions are expected to nearly coincide when both are rescaled to the same modulation depth. This is confirmed by the data shown in the inset of Figure \ref{fig:3}a, where the interference contrast is plotted as a function of the effective one-dimensional tunneling barrier height. For the $m_F=0$ states, the situation is different: Since in this case the optical potential is not additionally modulated, resonant tunneling takes place to the three {\it nearest neighbors}. This leads to a strongly increased tunneling probability and therefore higher lattice beam powers are required for the SF--MI transition to occur. Interestingly, all three transitions nearly coincide when normalized to the height of the barrier, regardless of the effect of different hopping distances and possible number of accessible sites.
This discrepancy can be explained when considering in addition to {\it nearest neighbor} tunneling processes also {\it second-nearest neighbor} processes. This is confirmed in calculations using the Gutzwiller approach. Results of such calculations are shown in Fig.~\ref{fig:3}b for the experimental configurations with filling factors ranging between two and four per lattice unit cell. The numerical results show a good qualitative agreement with the observed experimental data.
\\[0.4cm]
The fact that the SF--MI transition occurs at different lattice beam intensities for different values of $\left|m_F\right|$  adds an exciting new dimension to our lattice geometry:  it opens the possibility to create spin-mixtures where a fully localized component is immersed in a superfluid bath.
Experimentally, we study this configuration by loading $\ket{1,-1}$ atoms which preferably occupy the $\sigma^-$ lattice and add to these the more localized state $\ket{2,-2}$ ($\ket{2,+2}$),
which is strongly confined to the $\sigma^+$ ($\sigma^-$) lattice. As Figure~\ref{fig:3}c shows for the case of a 50:50 mixture, a considerable shift in the SF--MI transition of the $\ket{1,-1}$ atoms is observed when the admixed atoms occupy $\sigma^+$ sites, i.e., when they are located in between the preferably occupied $\sigma^{-}$ sites of $\ket{1,-1}$. On the other hand,  the transition remains practically unaltered when the more localized component occupies the same lattice sites as the superfluid core. The pictograms in Figure~\ref{fig:3}c illustrate these three different cases. The addition of a spin-crystal thus allows one to tailor the position of the SF--MI transition depending on the sub-lattice it occupies.
\\
The Gutzwiller computations reproduce the shift of the SF--MI transition (see Figure \ref{fig:3}d). The observed changes in the tunneling probabilities can be ascribed to a modification of second order hopping processes:
Second order hopping of $\ket{1,-1}$ between the $\sigma^-$ sites is only possible by passing through sites of the $\sigma^+$ lattice. However, in a $\ket{1,-1}$ -- $\ket{2,-2}$ mixture, these sites are occupied by $\ket{2,-2}$ atoms. The interaction energy between the atoms increases the effective barrier and suppresses tunneling. Thus, the SF--MI transition is shifted towards lower lattice laser beam intensities. By changing the relative composition of the mixture, the strength of this tunneling suppression can be modified.
\begin{figure*}
\includegraphics[width=0.7\textwidth]{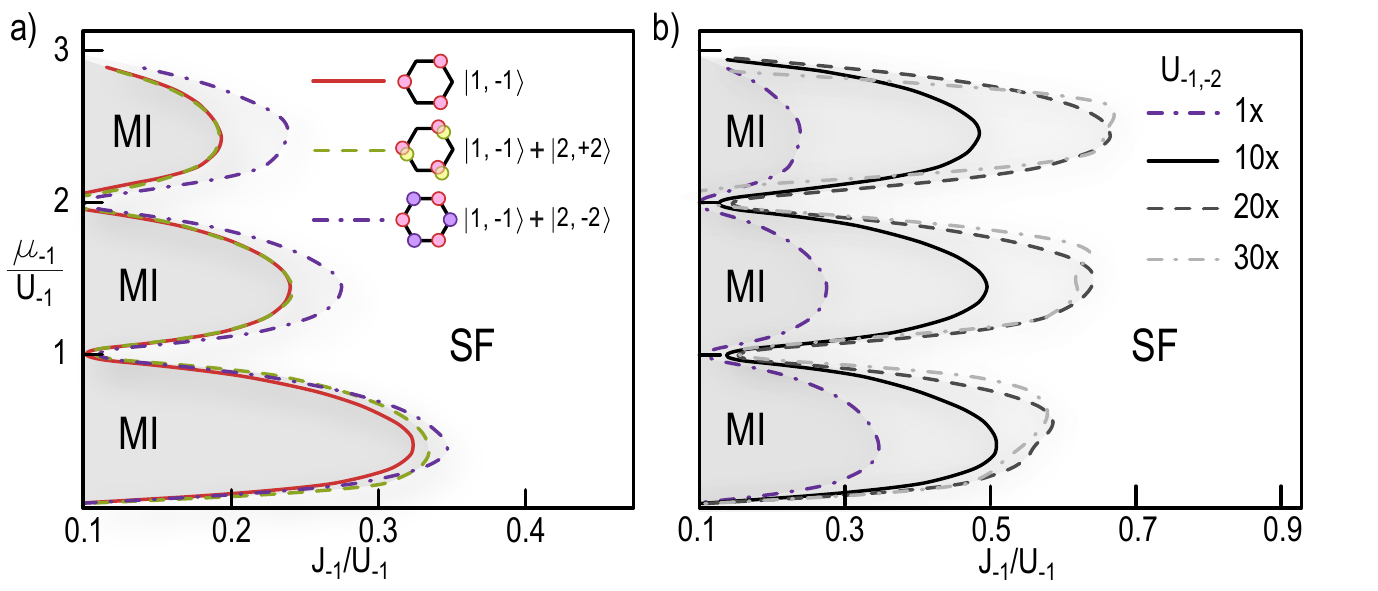}
    \caption{Superfluid to Mott-insulator transition phase diagram from Gutzwiller mean-field theory for $\ket{1,-1}$ atoms as a function of $J_{-1}/U^{\sigma^-}_{-1}$ and the chemical potential $\mu_{-1}/U^{\sigma^-}_{-1}$. \textbf{a} An admixture of $\ket{2,-2}$-atoms (blue dot-dashed line) strongly enlarges the $\ket{1,-1}$ Mott-lobes with respect to the pure system (red solid line) while an admixture of $\ket{2,+2}$-atoms (green dashed line) hardly affects the transition. \textbf{b} In a $\ket{1,-1}$--$\ket{2,-2}$-mixture the $\ket{1,-1}$ Mott-lobes grow when increasing the interaction strength between the spin-components (given in multiples of the $^{87}$Rb atomic interspecies interaction strength between $\ket{1,-1}$ and $\ket{2,-2}$).}
        \label{fig:4}%
\end{figure*}
\\
Further insight into the underlying physical processes is gained by considering the phase diagrams calculated in Gutzwiller approximation. Figure \ref{fig:4}a shows the SF--MI phase diagram for the experimental configurations discussed in Figs.~\ref{fig:3}c and d. A clear modification of the Mott-insulator lobes of a {\it triangular} state can be observed when strongly localized atoms are present in the complementary triangular sublattice($\ket{1,-1}+\ket{2,-2}$). Such a strong modification with respect to the pure case does not occur when both components occupy the same sites ($\ket{1,-1}+\ket{2,+2}$). This perfectly agrees with our experimental observations discussed above. In Fig.~\ref{fig:4}b we plot the SF--MI transition for a $\ket{1,-1}-\ket{2,-2}$ mixture for different inter-component interaction strengths $U_{-1,-2}$ between $\ket{1,-1}$ and $\ket{2,-2}$ ranging from the experimental value $U_{-1,-2}$ to $30\,U_{-1,-2}$. Such a situation of e.g.~$30\,U_{-1,-2}$ could be achieved experimentally by tuning the interactions with the help of Feshbach resonances\cite{Inouye1998}. Interestingly, the standard tendency is inverted in the hard core limit, when different components occupy different sites. This implies that the SF--MI transition for higher Mott orders occurs in the interesting regime that interpolates between the Bose--Hubbard regime and an array of coupled rotors, similar to an array of Josephson junctions\cite{Doniach1981,Sachdev}.
\section{Conclusion}
In conclusion, we have presented a comprehensive experimental and theoretical study of a novel type of optical lattice, which exhibits hexagonal spatial symmetry. The intriguing polarization structure of the lattice sites can cause a break up of the hexagonal structure for particular spin-states leading  to the creation of a triangular, magnetic sublattice.
The nature of this lattice potential, combined with the possibility to precisely control system parameters, makes this system especially suitable for studying new quantum phase mixtures of superfluid and Mott-insulating states.\\
We could show that a combined effect of repulsive interaction and the specific hexagonal lattice structure leads to modifications of $J/U$, which is clearly a different situation compared to e.g.~Bose-Fermi mixture experiments\cite{Gunter2006,Ospelkaus2006} where higher order effects effectively modify $J$ and $U$. \\
As promising further directions, entropy effects and entropic cooling \cite{Catani2009} can be studied in the hexagonal system. Particularly interesting, however, is the prospect of studying polarized or unpolarized Fermi gases,  or Fermi--Bose mixtures in the hexagonal lattice geometry. At half filling, both of these polarized systems will mimic the physics of graphene\cite{Lee2009}, i.e., exhibit Dirac's dispersion relation, artificial relativistic effects, etc. Introducing artificial gauge fields in such situations\cite{Lin2009,Lin2009b} will lead to the occurrence of the anomalous quantum Hall effect and a whole variety of exotic quantum phase transitions (see for instance Ref.~\cite{Bermudez2009}).
Furthermore, by only changing the polarization of the laser beams creating the lattice, a spin-independent triangular lattice can be created\cite{Becker2009}. A bosonic gas in the triangular lattice might then be used to mimic frustrated antiferromagnetism by employing a time-dependent lattice modulation \cite{Eckardt2010}.
\begin{acknowledgments}
We thank Dirk-S\"oren L\"uhmann and Andr\'e Eckardt for stimulating discussions. Moreover, we thank the Deutsche Forschungsgemeinschaft DFG for financial support within the Forschergruppe
FOR801 and the GRK 1355 and support by the Joachim Herz Stiftung. Support by the spanish MINCIN (FIS2008-00784 and Consolider QOIT), the Alexander von Humboldt foundation, Caixa Manresa, ERC grant QUAGATUA, and by the EU STREP NAMEQUAM are gratefully acknowledged.
\end{acknowledgments}
\newpage
\appendix
\section*{Appendix}
\noindent
{\bf Spin-dependent hexagonal optical lattice. }The light for the optical lattice is derived from a Ti:sapphire laser running at $\lambda=830$\,nm. To realize the two-dimensional hexagonal lattice we interfere three laser-beams under an angle of $120^\circ$ in one plane. The polarizations of the beams are linear and in the plane of the intersecting area and perpendicular to a homogeneous magnetic guiding field of $B_0=1.1\mathrm{G}$ which defines the quantization-axis. In order to compensate for phase-noise, the phase of each laser-beam is actively stabilized. To ensure two-dimensionality of the hexagonal structure, a perpendicular one-dimensional lattice with different frequency is applied.  Typical potential depths of this 1D-lattice are $V_{1\mathrm{D}}=44 E_{\mathrm{rec}}$. The resulting two-dimensional, hexagonal lattice potential is given by
\begin{align} \label{eq:Vlat}
  V(\bx) &= -2 V_\mathrm{lat} \Big \{ 3 - \cos\left [(\bb_1-\bb_2)\cdot \bx\right] + \cos\left (\bb_1 \cdot \bx\right) \nonumber \\
  & + \cos\left(\bb_2 \cdot \bx\right) \! \Big \}  - 2 \sqrt{3} V_\mathrm{lat} \alpha \, m_F \Big \{ \sin\left [(\bb_1-\bb_2)\cdot \bx\right] \nonumber \\
  & + \sin\left (\bb_1 \cdot \bx\right) - \sin\left(\bb_2 \cdot \bx\right)  \! \Big \} \nonumber \\ 
  & \equiv V_{\mathrm{hex}}\left(\bx\right) + m_F g_F \mu_{\mathrm{B}} B_{\mathrm{eff}}\left(\bx\right), 
\end{align}
where $\bb_1=\sqrt{3}/2 k \be_x + k/2 \be_y$ and $\bb_2=\sqrt{3}k \be_x$ with $k=2\pi/\lambda$ are the reciprocal vectors of the hexagonal lattice. The constant $\alpha$ depends on the detuning of the lattice laser from the $^{87}$Rb resonances at 780/795nm and is given for the experiments reported here ($\lambda=830\,\mathrm{nm}$) $\alpha=0.13$. The total potential is then given by $V(\bx) + V_{1\mathrm{D}} \sin^2(kz)$. The one-dimensional representation (Figure \ref{fig:1}b) can be reduced to a sinusoidal potential with 319\,nm period, modulated by a cosine function with 638\,nm period. The energy structure of the system can be readily calculated using an {\it ab initio} single-particle 2D band-structure calculation. \\[0.4cm] 
{\bf Preparation scheme. }We start with a Bose-Einstein condensate of typically several $10^5$ atoms in the stretched state $\ket{1,-1}$, which are confined in a nearly isotropic crossed dipole trap $(\omega \approx 2\pi \times 90 \mathrm{Hz}$). The preparation of the different pure and mixed spin-states is performed with aid of radio-frequency and/or microwave sweeps. After the state preparation we ramp up the optical lattice within 80ms using an exponential ramp. Within the ramping time the coherence between different spin-states is lost. \\[0.4cm] 
{\bf Spectroscopy. }We apply a microwave square-pulse of typically 2\,ms duration allowing for a sufficient resolution in frequency space. In order to minimize finite-state interactions, the intensity of the microwave pulse is adjusted such that typically only a small fraction of up to 20\% of the atoms populate the final state. To separate different spin-components during 27\,ms time-of-flight prior to absorption imaging, a Stern-Gerlach gradient field is applied. \\
\\[0.3cm]
{\bf Bose--Hubbard model and Gutzwiller mean-field Ansatz. }For the computation of the theoretical phase diagrams (Figs.~\ref{fig:3}b,~d, and Fig.~\ref{fig:4}) the system is modeled by a Bose--Hubbard model, which is known to give an essentially exact description of ultra-cold atoms in optical lattices\cite{Jaksch1998}. The corresponding Hamiltonian reads
\begin{align} \label{eq:H}
    H & = \sum_{\alpha=a,b}\!\!\Big\{-J_{\alpha} \sum_{\langle ij\rangle} \big( {\hat \alpha}_i^{\dagger} {\hat \alpha}_j + \textrm{h.c.} \big) \nonumber \\
      & + \frac{1}{2} \!\!\!\!\! \sum_{\substack{ i\in \sigma \\ \sigma=\sigma^+,\sigma^-}} \!\!\!\!\! \Big ( U^\sigma_{\alpha} \Big[ \hat n_{{\alpha},i} \Big( \hat n_{{\alpha},i} -1\big) 
       -\Big(\mu_{\alpha}-E^\sigma_{\alpha} \Big) \hat n_{{\alpha},i} \Big] \nonumber \\
      &+ U^{\sigma}_{a,b} \,\hat n_{a,i} \hat n_{b,i} \Big ) \Big\},
\end{align}
where $a$ and $b$ denote the two different spin-components and $\sigma^+$ and $\sigma^-$
distinguish between the two sublattices. The operator $\hat a_i$ ($\hat b_i$) destroys an $a$ ($b$) boson at site $i$, and $\hat n_{a,i}$ ($\hat n_{b,i}$) is the corresponding occupation operator. Angle brackets denote pairs of {\it nearest neighbors}. The tunnel matrix element is denoted by $J_{\alpha}$, the on-site interaction between the spin-components $\ket{\alpha}$ and $\ket{\alpha'}$ at $\sigma^{\pm}$ site by $U^{\sigma^{\pm}}_{\alpha,\alpha'}(U^{\sigma^{\pm}}_{\alpha}\!\!\! \equiv \!\!U^{\sigma^{\pm}}_{\alpha,\alpha})$. Local potentials are denoted by $E^\sigma_{\alpha}$ and the chemical potentials by $\mu_{\alpha}$. The parameters $E^\sigma_{\alpha}$, $U^{\sigma^{\pm}}_{\alpha,\alpha'}$,  and $J_{\alpha}$ are extracted from the experimental setting. For the derivation of the on-site interaction $U$ we integrate over the Gaussians which approximate well the Wannier functions at the minima of the optical lattice. We estimate the {\it nearest neighbor} tunneling $J$ from exact one-particle band-structure calculations of the full optical lattice. Comparing this to a tight-binding Hamiltonian gives the desired tunneling coefficients.\\
To solve the model Hamiltonian Eq.~\eqref{eq:H} we have employed the standard Gutzwiller mean-field approximation which is known to give good qualitative and quantitative account of the phase diagram in two and more dimensions and which reproduces the exact solution for $U/J\to\infty$ and, in the thermodynamic limit, also for $U/J\to0$.\cite{Lewenstein2007} The superfluid order parameter is defined as the expectation value of the annihilation operator. In order to calculate the phase diagrams in Fig.~\ref{fig:4}, we have chosen the chemical potential to be equal for both components, i.e., $\mu_{-1}=\mu_{\pm 2}$. In the $\ket{1,-1}-\ket{2,-2}$ mixture this leads to an approximately 50:50 mixture, while in the $\ket{1,-1}-\ket{2,+2}$ mixture the $\ket{1,-1}$ atoms repel the $\ket{2,+2}$ strongly such that the occupation number can be heavily asymmetric.
\end{document}